# Mnesors for databases

Gilles CHAMPENOIS

Collège Saint-André, Saint-Maur, France

gilles_champenois@yahoo.fr

ABSTRACT. We add commutativity to axioms defining mnesors and substitute a bitrop for the lattice. We show that it can be applied to relational database querying: set union, intersection and selection are redefined only from the mnesor addition and the granular multiplication. Union-compatibility is not required.

> *"If you tell me you have 50 different ways of representing data in your system at the logical level, then I'll tell you that you have 49 too many."*
>
> *Ted Codd (1923-2003)*

## I. INTRODUCTION

The theory of mnesors is here modified by assuming commutativity. We used for instance to consider the colomn tuple $\begin{bmatrix} Sweden \\ Germany \end{bmatrix}$ as an ordered stack, *Sweden* being older in the stack than *Germany*. So $\begin{bmatrix} Sweden \\ Germany \end{bmatrix}$ was different to $\begin{bmatrix} Germany \\ Sweden \end{bmatrix}$. Now by assuming commutativity ($x + y = y + x$), the order does not matter, i.e. $\begin{bmatrix} Sweden \\ Germany \end{bmatrix}$ equals $\begin{bmatrix} Germany \\ Sweden \end{bmatrix}$. So we will simply write $\begin{Bmatrix} Sweden \\ Germany \end{Bmatrix}$.

Our purpose is to find a mathematical model for information [1][2] similar to abstract vectors. We first define a structure called a bitrop which plays the role of the field for vectors, and then we will define semimodules over bitrops. The resulting two-sorted structure forms an interesting linear algebra because it mixes properties of lattices (boolean for example) with properties of vectors: the addition of mnesors aggregates information and the external multiplication selects information.

We first define what we call a bitrop. A bitrop $B$ is a set with two operations, a commutative addition and a distributive multiplication. Moreover, it has an element $\tau$ called the center of $B$ verifying the next two properties.

- $x \otimes \tau = x$    for any $x \in B$ (1)

- $\tau \oplus \tau = \tau$ (2)

- Two other properties are added. The subset of elements verifying $\lambda \oplus \tau = \tau$ is denoted by $B^+$. So for any $x, y \in B$ and $\lambda, \mu \in B^+$

    - there exists $\alpha \in B^+$ such that $(x \oplus y) \otimes \alpha = x$ (3) absorption property

    - $x \otimes \lambda = x \otimes \mu$ implies $\lambda = \mu$ (4) cancellation property

EXAMPLE. The min-plus integers with the minimum as addition ($+$) and the addition as multiplication ($\otimes$) form a commutative bitrop.

## II. DEFINITION OF A COMMUTATIVE MNESOR SPACE

A mnesor space is a semimodule ($M$) over a commutative bitrop. $M$ possesses an addition (written $+$) and an identity element $0$. According to vector scalars, bitrop elements are called granulars. The commutative semimodule $M$ verifies the following properties:

| | | |
|---|---|---|
| ☑ (unital property) | $x \tau = x$ | (4) |
| ☑ (mnesor distributivity) | $(x + y) \lambda = x \lambda + y \lambda$ | (5) |
| ☑ (associativity of granular multiplication) | $(x \lambda) \mu = x (\lambda \otimes \mu)$ | (6) |
| ☑ (granular distributivity) | $x (\lambda \oplus \mu) = x \lambda + x \mu$ | (7) |

The absorption property is added:

☑ For any mnesors $x, y$, there exists a granular $\alpha \in B^+$ such that $(x + y) \alpha = x$    (8)

*Idempotence*. The addition of mnesors is idempotent.

PROOF. Letting $\lambda = \mu = \tau$ in (5) yields $x\tau + x\tau = x(\tau \oplus \tau) = x\tau$. Thus, $x + x = x$, for any $x \in M$ [by (4)]

*Addition ordering*. A mnesor $x$ is a prefix of a given mnesor $a$ if we can write $x + a = a$. The relation "is a prefix of" is an order relation (since $M$ is a commutative idempotent monoid) called addition ordering.

*Multiplication ordering*. "$x$ is lower than $a$" iff we can write $x = a\lambda$ where $\lambda \in B^+$ ($x$ is said to belong to the orbit set of $a$).

PROOF. Reflexivity ($x = x\tau$),
 Antisymmetry ($x = a\lambda$ and $a = x\mu$ implies
 $x = x\tau = x(\tau \oplus \mu) = x\tau + x\mu = x + a = a\lambda + a = a$),
 Obviously transitivity.

The two orderings are equivalent, that is $x = a\lambda$ iff $x + a = a$.

PROOF.
($\leftarrow$) We apply property (8). There exists $\lambda \in B^+$ such that $(x + a)\lambda = x$. But $x + a = a$, then, $a\lambda = x$
($\rightarrow$) $x + a = a\lambda + a = a\lambda + a\tau = a(\lambda \oplus \tau) = a\tau = a$

*Mnesor intersection.* Let be $x, y \in M$ given. Since the multiplication is commutative, then there exist $\lambda, \mu \in B^+$ such that $x\lambda = y\mu$.

PROOF. $(x + y)\lambda = x$ and $(y + x)\mu = y$. By multiplying by $\mu$ and $\lambda$, we get $((x + y)\lambda)\mu = x\mu$ and $((y + x)\mu)\lambda = y\lambda$. Thus $(x + y)(\lambda \otimes \mu) = x\mu = y\lambda$.

If $\lambda'$ is another granular that verifies $(x + y)\lambda' = x$, then $y\lambda' = y\lambda = x\mu$. Thus, $y\lambda$ is uniquely defined for each $x$. $x\lambda = y\mu$ is denoted by the operation $x \mathbin{\text{O}} y$ (the intersection of $x$ and $y$). Note that the intersection is commutative, associative and idempotent.

*Lattice.* $(M, +, \text{O})$ is a lattice.

PROOF. The addition and the multiplication of mnesors are commutative, associative, idempotent. Absorption laws hold: $x + (x \mathbin{\text{O}} y) = x + x\mu = x\tau + x\mu = x(\tau \oplus \mu) = x\tau = x$ and $x \mathbin{\text{O}} (x + y) = x\tau = x$.

## III. EXAMPLES

Three of the operations of relational algebra can be expressed with mnesor addition and external multiplication. But here there's no need for union-compatibility.

*Set union.* The set union of tables $x$ and $y$ is performed by the addition $x + y$.

EXAMPLES. The bitrop is the boolean lattice of intergovernmental organization membership.
$$\begin{Bmatrix} Sweden \\ Germany \end{Bmatrix} + \begin{Bmatrix} France \\ Sweden \end{Bmatrix} = \begin{Bmatrix} Sweden \\ Germany \\ France \end{Bmatrix}$$
$$europe\, NATO + europe\, \overline{NATO} = europe\left(NATO \oplus \overline{NATO}\right) = europe$$

*Intersection*. The intersection of tables $x$ and $y$ is performed by $x \circ y$.

EXAMPLE. $\begin{Bmatrix} Germany \\ Denmark \end{Bmatrix} \circ \begin{Bmatrix} Germany \\ Sweden \end{Bmatrix} = \{Germany\}$ because there exists $\lambda = NATO$ such that

$$\left( \begin{Bmatrix} Germany \\ Denmark \end{Bmatrix} + \begin{Bmatrix} Germany \\ Sweden \end{Bmatrix} \right) \lambda = \begin{Bmatrix} Germany \\ Denmark \\ Sweden \end{Bmatrix} \lambda = \begin{Bmatrix} Germany \\ Denmark \end{Bmatrix}$$

Thus, $\begin{Bmatrix} Germany \\ Sweden \end{Bmatrix} \lambda = \{Germany\}$

*Selection*. It is directly performed by the external multiplication. Properties (6) of mnesors is equivalent to the composition of selections.
A case of filtering failure: $Sweden\ EU = Sweden$ and case of matching failure: $Australia\ EU = 0$.

EXAMPLE. The selection of countries with $\lambda = EU$-membership: $x \lambda$ where $x$ is the tuple representing all the countries wordwide.

## REFERENCES

1. KOHLAS J. (2003), *Information Algebras: Generic Structures for Inference,*
2. TROPASHKO V. (2005), *Relational Algebra as non-Distributive Lattice.*